\documentclass[letterpaper,nohyper]{JHEP3}

\usepackage{amssymb,amsmath, amsfonts}
\usepackage{graphicx}
\usepackage{subfigure}

\def\be{\begin{eqnarray}}
\def\ee{\end{eqnarray}}

\newcommand{\eqn}[1]{(\ref{#1})}

\title{Non-Abelian Berry Phase, Instantons and $\mathcal N = (0,4)$ Supersymmetry}
\author{Jo\~ao N. Laia\\
Department of Applied Mathematics and Theoretical Physics, \\
University of Cambridge, UK\\
{\tt j.laia@damtp.cam.ac.uk}}
\abstract{In supersymmetric quantum mechanics, the non-Abelian Berry phase is known to obey certain differential equations. Here we study ${\cal N}=(0,4)$ systems and show that the non-Abelian Berry connection over ${\bf R}^{4n}$ satisfies a generalization of the self-dual Yang-Mills equations. Upon dimensional reduction, these become the tt* equations. We further study the Berry connection in ${\cal N}=(4,4)$ theories and show that the curvature is covariantly constant.}


\begin{document}

\section{Introduction\label{intro}}

Although the concept of a non-Abelian gauge connection is most familiar in the context of
particle physics, the first place that one encounters this idea is in the more down-to-earth
framework of non-relativistic quantum mechanics. Suppose that a quantum system has an $N$-fold
degeneracy of states, $|a\rangle$, for $a=1,\ldots,N$. Moreover, we assume that this degeneracy
is preserved as one varies a set of external parameters, $\phi_p$. Then one can define a $U(N)$
gauge connection over the space of parameters,
\be (A_p)_{ab} = \langle b|\frac{\partial}{\partial\phi_p}|a\rangle\label{A} \;\; .\ee
Far from a mathematical curiosity, this gauge connection has physical consequences. As
one adiabatically varies the parameters $\phi_p$ around a closed path, a state
within this energy subspace undergoes a $U(N)$ holonomy which depends on the path taken,
\begin{equation}\label{aiai}
U = T \exp \left( i \int_{t_i}^{t_f} A_p (\phi) \dot \phi ^p dt  \right) \;\; .
\end{equation}

This is known as the non-Abelian Berry phase or, more precisely, Berry holonomy
 \cite{Berry:1984jv,Moody:1989vh}. However, in general the connection $A_p$ is difficult to
compute. Indeed, one can see from \eqn{A} that an explicit derivation requires
us to know the exact energy eigenstates $|a\rangle$ for all values of the
parameters $\phi_p$. Apart from in a few simple (but important!) systems,
this is information that we don't have access to.

Progress can be made if one considers supersymmetric quantum mechanics. In this case, one can show
that the Berry connection $A_p$ must satisfy certain differential equations. In certain cases,
this makes the problem of computing the connection tractable even when the ground states are
not explicitly known. Of course, the systems that we can study in a laboratory are not supersymmetric.
Nonetheless, it may still be possible to make progress by truncating supersymmetric systems to
more realistic non-supersymmetric quantum mechanics which inherit the Berry connection from their
supersymmetric parent -- an example of this was given in \cite{Sonner:2008be}.

The first example of a non-Abelian Berry phase in supersymmetric quantum mechanics was given by
Cecotti and Vafa \cite{Cecotti:1991me}. They studied theories with $\mathcal N = (2,2)$ supersymmetry.
When the parameters $\phi_p$ live in supersymmetric chiral multiplets, they showed that the
resulting Berry connection obeys the so-called tt* equations,
\begin{subequations}\label{ttstar}
\begin{align}
{\cal D}_p G_q &= {\cal D}_q G_p  \ \ \ \  ,\ \ \ \   [G_p^\dagger,G_q] = [{\cal D}_p,{\cal D}_q^\dagger] \;\; ,\\
{\cal D}_p^\dagger G_q &= [G_p,G_q] = [{\cal D}_p,{\cal D}_q]=0 \;\;.\label{13b}
\end{align}
\end{subequations}
Here ${\cal D}_p$ is the covariant derivative with respect to the complex parameter $\phi_p$:
${\cal D}_p\equiv \partial/\partial\phi_p + i[A_p,.]$. Meanwhile $G_p$ is a particular correlation
function within the theory that may typically be computed semi-classically for large values of $\phi$.

A simple derivation of the tt* equations was presented recently in \cite{Sonner:2008fi}.
One generalizes the connection $A_p$ in \eqn{aiai} to a supersymmetric connection, depending both on
bosonic parameters $\phi_p$ and their fermionic partners.
\be {\cal A} = A_p(\phi)\dot{\phi}_p+\ldots \;\; ,\ee
where $\ldots$ include the fermions.
In order for the system to be supersymmetric, the holonomy \eqref{aiai} should remain invariant up to a relabelling of the basis vectors \cite{Herbst:2008jq, Sonner:2008fi}. We therefore require that under supersymmetry, the connection transforms as,
\begin{equation}\label{sym}
\delta {\cal A} = \frac{d\Theta}{dt} + i [{\cal A}, \Theta ] \;\; .
\end{equation}
To illustrate this technique, if we take the parameters of the system to live in chiral multiplets, the most general $U(N)$ connection we can write is
\begin{equation}\label{chirallagrangian}
{\cal A} = (A_p  \dot \phi^p + G_p F^p - B_{pq} \psi^p \psi^q + \text{h.c.} ) + C_{pq} \bar \psi^p \psi^q + \vec C_{pq} \cdot \bar \psi^p \vec\sigma \psi^q \;\; ,
\end{equation}
where $A_p$, $G_p$ and $B_{pq}$ are $u(N)$-valued matrices, which  are complex functions of $\phi$ and $\phi^\dagger$, while $C_{pq}$ and $\vec C_{pq}$ are real $u(N)$-valued functions of $\phi$ and $\phi^\dagger$. Imposing the transformation \eqn{sym} requires that the connections 
$A_p$ and the matrices  $G_p$ obey the tt* equations \eqn{ttstar} while the remaining matrices are given by
\begin{subequations}\label{chiralconditions}
\begin{align}
\vec C_{pq} &= 0  \ \ \ \  ,\ \ \ \    B_{pq} + B_{qp}=\mathcal D_p G_q \;\; ,\\
C_{pq} &= [G_p^\dagger, G_q] = [{\cal D}_p,{\cal D}_q^\dagger] \;\;. \label{17b}
\end{align}
\end{subequations}

One advantage of this new technique is that it allows us to find the possible Berry connections obeyed
by more general parameters. For example, an ${\cal N}=(2,2)$ vector multiplet contains a triplet of
parameters (as opposed to the complex parameter in the chiral multiplet). It was shown in \cite{Sonner:2008fi} that
the non-Abelian Berry connection must now satisfy the Bogomolnyi monopole equations over ${\bf R}^3$
or, for multiple parameters, a novel generalization of the Bogomolnyi equation over ${\bf R}^{3n}$, with $n$ being the number of vector multiplets.

The purpose of this paper is to extend this technique to further supersymmetric multiplets 
 to determine the associated Berry connections. In section \ref{sec2}, we consider parameters in a $\mathcal N = (0,4)$ multiplet. We will show that the Berry connection in such systems is forced to obey the self-dual Yang-Mills equations. We further show that when one varies parameters in $n$ multiplets, the Berry connection is given in terms of a novel generalization of the self-dual equations to ${\bf R}^{4n}$. In section \ref{sec3}, we derive the constraints on the Berry connection imposed by $\mathcal N = (4,4)$. We use that information to reproduce an earlier result \cite{deBoer:2008ss} where a different approach was used.

\section{$\mathcal N = (0,4)$ supersymmetry and the self-dual equations\label{sec2}}

In this section, we generalise the Berry's phases studied in \cite{Sonner:2008fi} by considering different supersymmetries. Our focus is on $\mathcal N = (0,4)$ supersymmetry with supersymmetry transformations given by
\begin{subequations}\label{epanaosei}
\begin{align}
\delta X^a_r &= i \beta_4 \psi^a_r +i \beta_m \psi^b_r \mathcal J_b^{(m)\,a} \;\; , \\
\delta \psi^a_r &= -\dot X^a_r \beta_4 + \beta_m \dot X^b_r \mathcal J_b^{(m)\,a} \;\; .
\end{align}
\end{subequations}
Our conventions are those of \cite{Gauntlett:1993sh}. Here, $r=1,\ldots, n$ labels the multiplets, while $a=1,2,3,4$ is an R-symmetry index. The $X_r$ are real bosonic variables, $\psi_r$ real fermionic variables, $\beta$ the four Grassmann-valued supersymmetric parameters, and $J_b^{(m)\,a}$, $m=1,2,3$ are almost complex structures.
The most general connection we can have is
\begin{equation}\label{generalLagrangian}
{\cal A} = A_a^r (X) \dot X^a_r + B_{ab}^{rq} (X) \psi^a_r \psi^b_q \;\; ,
\end{equation}
with $A_a^r$ and $B_{ab}^{rq}$ matrices. Also, we can see that $B$ satisfies the antisymmetry condition $B_{ab}^{rq}=-B_{ba}^{qr}$. After a straightforward computation, which we 
reproduce in the appendix, one finds that ${\cal A}$ is invariant under the supersymmmetry transformations \eqref{epanaosei} if the fields obey
\begin{subequations}
\begin{align}
&F_{ab}^{rq} = 2i B_{ab}^{rq} \;\; ,\label{minst0}\\
&F_{ab}^{rq} \mathcal J^{(m)\,b}_c= 2 i B_{cb}^{qr}\mathcal J^{(m)\,b}_a\;\; , \label{minst}\\
& \left( \mathcal D_c^s B_{[ab}^{rq} \right) \mathcal J^{(m)\,c}_{d]} = 0 \;\; , \label{dontknow}
\end{align}
\end{subequations}
The field strength is defined in the usual manner: $F_{ab}^{rq} = \partial A_b^q / \partial X^a_r - \partial A_a^r / \partial X^b_q + i [A^r_a , A^q_b]$. In the last of these equations, the anti-symmetrisation over the indices $abd$ also rotates  the indices $rqs$. Equations \eqref{minst0} and \eqref{minst} imply a 
restriction on the connection, 
\begin{equation}\label{inst}
F^{qr}_{ca}\mathcal J^{(m)\,c}_{b} - F^{rq}_{cb}\mathcal J^{(m)\,c}_{a}= 0 \;\; ,
\end{equation}
For the case of a single multiplet (i.e. $n=1$), these equations were previously derived in \cite{Howe:1988cj}. The case of many multiplets appears to be novel.

Equation \eqref{dontknow} can be written in a slightly different form, by using the Bianchi identities and equation \eqref{inst},
\begin{equation}
F_{c[a}^{sr} \partial_b^q \mathcal J^{(m)\; c}_{d]} = 0
\end{equation}
where, again, the anti-symmetrisation over the indices $abd$ also rotates  the indices $rqs$.
A trivial solution to this equation is achieved if one considers almost complex structures in ${\bf R}^{4}$, given by the 't Hooft matrices
\begin{equation}\label{hooft}
\mathcal J^{(1)} = \left( 	\begin{array}{cccc}
0 & 0 & 0 & 1\\
0 & 0 & 1 & 0\\
0 & -1 & 0 & 0\\
-1 & 0 & 0 & 0
\end{array} \right) \quad , \quad
\mathcal J^{(2)}  = \left( 	\begin{array}{cccc}
0 & 0 & -1 & 0\\
0 & 0 & 0 & 1\\
1 & 0 & 0 & 0\\
0 & -1 & 0 & 0\end{array} \right) \quad , \quad
\mathcal J^{(3)}  = \left( 	\begin{array}{cccc}
0 & 1 & 0 & 0\\
-1 & 0 & 0 & 0\\
0 & 0 & 0 & 1\\
0 & 0 & -1 & 0\end{array} \right) \;\; .
\end{equation}
Equation \eqref{inst} then becomes a generalization of the self-dual equations for a 
$U(N)$ connection over ${\bf R}^{4n}$,
\begin{equation}\label{ginst}
F_{\mu\nu}^{rq} = -\frac{1}{2} \epsilon_{\mu\nu\rho\sigma} F_{\rho\sigma}^{rq} - \delta_{\mu\nu} Z^{rq} \;\; ,
\end{equation}
where $Z^{rq}=-F_{11}^{rq}=-F_{22}^{rq}=-F_{33}^{rq}=-F_{44}^{rq}$ is antisymmetric in $r \leftrightarrow q$. Obviously, if we restrict to a single multiplet, these equations 
reduce to the familiar instanton equations on ${\bf R}^4$. In the appendix, we rederive 
\eqn{ginst} as the restriction on a Berry connection starting from a different (and seemingly less standard) set of supersymmetry transformations. 

The above results show that if we vary parameters of supersymmetric quantum mechanics, living in an ${\cal N}=(0,4)$ multiplet with transformations \eqn{epanaosei}, the associated
Berry phase must obey \eqn{inst}. It remains to find an explicit example. Moreover, there appear to be some conceptual difficulties in constructing such a system. In the case of the 't Hooft 
Polyakov monopole, the off-diagonal terms in the connection could be understood in terms of tunnelling processes in quantum mechanics \cite{Sonner:2008be} whose characteristic feature is an exponential fall-off. The interpretation of the instanton power-law fall-off is presently unclear.

\subsection{Dimensionally reducing the self-dual equations}

It is known that dimensional reduction of the self-dual equations leads to the Bogomolnyi equation. Also, we know that a second dimensional reduction allow us to obtain the Hitchin equation. In this subsection, we show that there are similar relations between the generalisation of the self-dual equations \eqref{ginst}, the tt*-Bogomolnyi equations of \cite{Sonner:2008fi} and the tt* equations \eqn{ttstar}.
We start by dimensionally reducing \eqref{ginst} along the directions $X^4$ of every multiplet, labelled by $q$. Define $H_q \equiv - A_4^q$. The dimensional reduction of the field strength tensor gives
\begin{subequations}
\begin{align}
F_{44}^{qp} &\equiv i [H_q,H_p] = -Z^{qp} = F_{ii}^{qp} \qquad \text{(no sum over $i$)} \;\; , \label{dimred1} \\
F_{i4}^{qp} &= F_{i4}^{pq} \quad \Leftrightarrow \quad \mathcal D_i^q H_p = \mathcal D_i^p H_q \;\; , \label{dimred2}\\
F_{ij}^{qp} &= - \epsilon_{ijk4}F_{k4}^{qp} +\delta_{ij} F_{44}^{qp} = \epsilon_{ijk} \mathcal D_k^q H_p + i \delta_{ij}  [H_q,H_p] \;\; . \label{dimred3}
\end{align}
\end{subequations}
These are the tt*-Bogomolnyi equations.

We now further dimensionally reduce along the directions $X^3$. Defining 
\begin{subequations}
\begin{align}
 G_p =& -A_4^p + i A_3^p  \;\;,\\
 A^p=& A_1^p - i A_2^p \;\;,\\
 \phi_p =& X^1_p + i X^2_p \;\;.
\end{align}
\end{subequations}
allows us to obtain the equations
\begin{subequations}
\begin{align}
[G_p,G_q] &= [\mathcal D_p, \mathcal D_q ] = 0 \;\; , \\
\mathcal D_p^\dagger G_q &= 0 \;\; ,\\
\mathcal D_p G_q &= \mathcal D_q G_p \;\; , \\
[\mathcal D_p, \mathcal D_q^\dagger ] &= [G_q^\dagger, G_p]\;\; .
\end{align}
\end{subequations}
But these are the tt* equations. %

\section{$\mathcal N = (4,4)$ supersymmetry \label{sec3}}

In this section, we consider the parameters to be living in an hypermultiplet of $\mathcal N = (4,4)$ supersymmetry. Its field content is the same as two $\mathcal N = (2,2)$ chiral multiplets,
\begin{subequations}\label{rel}
\begin{align}
(\phi_1, \psi_\pm, F_1) &\qquad \text{chiral multiplet 1} \;\; , \\
(\phi_2, \lambda_\pm, F_2) &\qquad \text{chiral multiplet 2} \;\; .
\end{align}
\end{subequations}
For consistency in notation, $\psi$ should be called $\psi_1$ and $\lambda$ should be $\psi_2$, but that could generate unnecessary confusions later on.

The Berry phase in this system will be given by the tt* equations with two chiral multiplets, subject to extra conditions coming from the extra supersymmetry that we now have. These extra symmetry can be looked at as an R-symmetry that rotates the chiral multiplets \eqref{rel} into each other. In this section, we are going to compute these conditions.

The R-symmetry of $\mathcal N = (4,4)$ supersymmetry is $SO(5)\times SU(2)$. $(\phi_1, \phi_2^\dagger)$ transforms in the $\bf{(1,2)}$ representation, i.e., a singlet under $SO(5)$ and a doublet under $SU(2)$. The 4 complex fermionic variables transform in the $\bf{(4,1)}$, while the $F$'s are singlets under both transformations. 

The restrictions from the $SU(2)$ part of the R-symmetry put $A_1$ and $A_2^\dagger$ in a doublet, while the other coefficients are singlets under it. In the rest of the section we focus on the conditions imposed by the $SO(5)\cong Spin(5)$ part.

In order to do that, it is useful to work in the chiral representation of the Clifford algebra, defined by having a charge matrix of the form %
\begin{equation}
B \propto  \left( 	\begin{array}{cc}
0 & i\sigma^2\\
i\sigma^2 & 0
\end{array} \right) \;\; ,
\end{equation}
In such representation, the $Spin(5)$ spinor can be written in terms of the two $Spin(2)$ spinors by putting one of them in the first two components, and the second one in the bottom ones,
\begin{equation}
\Psi = \left( 	\begin{array}{c}
\lambda_\alpha \\
\bar \psi^{\dot \alpha}
\end{array} \right) \;\; .
\end{equation}
The gamma matrices are given by
\begin{equation}
\gamma^i = \left( 	\begin{array}{cc}
\sigma^i & 0\\
0 & - \sigma^i
\end{array} \right) \qquad , \qquad
\gamma^4 = \left( 	\begin{array}{cc}
0 & i  1\\
-i 1 & 0
\end{array} \right) \qquad , \qquad
\gamma^5 = \left( 	\begin{array}{cc}
\; 0\; & \; 1\;\\
\; 1\; & \; 0\;
\end{array} \right) \;\; .
\end{equation}

We now write $SO(5)$ invariant interactions in terms of the $Spin(2)$ spinors. By comparison with \eqref{chirallagrangian}, we will be able to tell the extra conditions that the coefficients in the Berry phase have to satisfy. $SO(5)$ covariant interactions with two spinors are \cite{Polchinski:1998rr}
\begin{align}
\Psi^\dagger \Psi &\qquad SO(5) \text{ scalar} \;\; ,\\
\Psi^\dagger \gamma^\mu \Psi &\qquad SO(5) \text{ vector} \;\; ,\\
\Psi^\dagger \gamma^{[\mu} \gamma^{\nu]} \Psi &\qquad SO(5) \text{ 2-tensor} \;\; .
\end{align}
Writing them in terms of the $Spin(2)$ spinors,
\begin{subequations}
\be\label{duas}\Psi^\dagger \Psi = \bar \lambda \lambda - \bar \psi \psi \;\; ,\ee
\be\label{uma}
V_\mu  \Psi^\dagger \gamma^\mu \Psi = V_i (\bar \lambda \sigma^i \lambda - \bar \psi \sigma^i \psi ) + (V_5 + i V_4)\bar \lambda \bar \psi + (V_5 - i V_4) \lambda \psi \;\; ,\ee
\be\begin{split}\label{tres}
V_{\mu\nu} \Psi^\dagger i\gamma^{[\mu} \gamma^{\nu]} \Psi = - \epsilon_{ijk} V_{ij} &(\bar \lambda \sigma^k \lambda + \bar \psi \sigma^k \psi )-2 V_{45} (\bar \lambda \lambda + \bar \psi \psi) \\
- 2(V_{i4} + i V_{i5})\psi^\alpha &(\sigma^i)_{\alpha}^{\;\beta} \lambda_\beta - 2(V_{i4} - i V_{i5})\bar \lambda_{\dot\alpha} (\sigma^i)^{\dot\alpha}_{\;\dot\beta} \bar \psi^{\dot\beta} \;\; .
\end{split}\ee
\end{subequations}

In the language of two chiral multiplets, we know that $\vec C_{pq} = 0$. That means that $V_{ij}=V_{i}=0$. But because we now have more symmetry, those coefficients are actually sitting on $SO(5)$ tensor and vector, respectively. Hence, by rotational symmetry we conclude $V_{\mu\nu}=V_{\mu}=0$. Therefore, interactions such as $\lambda \psi$ are not allowed in the Berry phase, giving the extra condition of $B_{12} = B_{21} = 0$

Finally, by looking at the interactions coming from $\Psi^\dagger \Psi$, we can say in the old language that $C_{11} = -C_{22}$. Also, since there are no interactions in equations \eqref{uma}, \eqref{duas} and \eqref{tres} of the type $\bar \lambda \psi$, $\lambda \lambda$ nor $\psi \psi$, we conclude that $C_{12} = C_{21} = 0$ and $B_{11} = B_{22} = 0$.

In sum, an effective description of a system with parameters living in a hypermultiplet of $\mathcal N = (4,4)$ supersymmetry will give rise to a Berry's phase with coefficients in \eqref{chirallagrangian} subject to the restrictions \eqref{ttstar}, \eqref{chiralconditions} and 
\begin{subequations}\label{44rest}
\begin{align}
\mathcal D_p G_q &= B_{pq}+B_{qp} = 0 \;\; , \label{fff2}\\
C_{pq} &= [G_p^\dagger, G_q] = [\mathcal D_p, \mathcal D_q^\dagger] = 0 \qquad , \qquad \text{for $p\neq q$}\;\; , \label{fff4}
\end{align}
\end{subequations}
where $p,q \in \{1,2\}$. 

The coefficient $C_{pq}$, by virtue of equation \eqref{17b}, is to be seen as a curvature. Restrictions \eqref{13b} and \eqref{fff2} imply $\mathcal D_p C_{rs} = 0$, meaning that $\mathcal N = (4,4)$ systems have a covariantly constant curvature.

This result has already been derived in the context of two-dimensional ${\cal N}=(4,4)$ superconformal field theories, where the bundle of chiral primary operators is shown to have covariantly constant curvature \cite{deBoer:2008ss}.

\section{Acknowledgements}
The author would like to thank David Tong for important discussions during every stage of this project.
The author acknowledges the support from the Funda\c c\~ao para a Ci\^encia e Tecnologia (FCT-Portugal) through the grant SFRH/BD/36290/2007.

\appendix
\section{Computation of the generalisation of the self-dual equations\label{apendice}}

\subsection{Using $\mathcal N = (0,4)$ supersymmetric multiplets}
For multiplets transforming under \eqref{epanaosei}, the most general connection is, by dimensional reasoning, given by equation \eqref{generalLagrangian}. Doing variations to it, we get
\begin{equation}
\begin{split}
\delta {\cal A} = & \frac{d}{dt}\left( A_a^r \delta X^a_r \right) + \partial_b^q A_a^r \delta X_q^b \dot X ^a_r - \partial_b^q A_a^r \delta X_r^a \dot X ^b_q \\ & + \partial_c^s B_{ab}^{rq} \delta X_s^c \psi_r^a \psi_q^b + B_{ab}^{rq} \delta \psi_r^a \psi _q^b + B_{ab}^{rq}  \psi_r^a \delta \psi _q^b \;\;,
\end{split}
\end{equation}
where $\partial_a^r = \partial / \partial X_r^a$. On the other hand, in order for this variation to be a symmetry, equation \eqref{sym} must hold, with $\Theta$ chosen to be $A_a^r \delta X^a_r$:
\begin{equation}
\delta {\cal A} = \frac{d}{dt}\left( A_a^r \delta X^a_r \right) + i\left[ A_a^r , A_b^s \right] \dot X ^a_r \delta X^b_s + i\left[ B_{ab}^{rq},A_c^s \right]\psi_r^a \psi_q^b \delta_s^c \;\; .
\end{equation}

Using the supersymmetry transformations, these two equations are telling us
\begin{equation}
\begin{split}
&\left( \partial_b^q A_a^r - \partial_a^r A_b^q   \right)\dot X_r^a \left( i \beta_4 \psi^b_q +i \beta_m \psi^c_q \mathcal J_c^{(m)\,b} \right) + \partial_c^s B_{ab}^{rq} \psi_r^a \psi_q^b \left(  i \beta_4 \psi^c_s +i \beta_m \psi^d_s \mathcal J_d^{(m)\,c}\right) \\
&\qquad \qquad  + B_{ab}^{rq} \left( -\dot X^a_r \beta_4 + \beta_m \dot X^c_r \mathcal J_c^{(m)\,a} \right) \psi^b_q + B_{ab}^{rq} \psi^a_r \left( -\dot X^b_q \beta_4 + \beta_m \dot X^c_q \mathcal J_c^{(m)\,b} \right) \\
=& i \left[ A_a^r , A_b^q \right] \dot X_r^a \left( i \beta_4 \psi^b_q +i \beta_m \psi^c_q \mathcal J_c^{(m)\,b} \right) + i \left[ B_{ab}^{rq} , A_c^s \right] \psi_r^a \psi_q^b \left( i \beta_4 \psi^c_s +i \beta_m \psi^d_s \mathcal J_d^{(m)\,c} \right) \;\; .
\end{split}
\end{equation}

The coefficients of $\beta_4 \dot X \psi$ and $\beta_m \dot X \psi$ give us the conditions \eqref{minst0} and \eqref{minst} respectively. The coefficient of $\beta_4 \psi \psi \psi$ vanishes due to Bianchi identity, whereas from the coefficient of $\beta_m \psi \psi \psi$ we get \eqref{dontknow}.

\subsection{Using another multiplet}
With the idea of dimensionally reducing the generalisation of the self-dual equation to get the tt* ones in mind, we can try to do something analogous at the level of the supersymmetry transformations. Consider the following transformations
\begin{subequations}\label{susytrans}
\begin{align}
\delta x^\mu_q &= i \bar \lambda^q \bar \sigma^\mu \xi - i \bar \xi \sigma^\mu \lambda^q \;\; ,\\
\delta \lambda^q &= \dot x^\mu_q \bar \sigma^\mu \xi \;\; , \label{eqeqeq}
\end{align}
\end{subequations}
where $\left(\bar \sigma^\mu\right)^{\alpha\beta} = (\sigma^i, i1)^{\alpha\beta}$ and  $\left(\sigma^\mu\right)_{\alpha\beta} = (\sigma^i, -i1)_{\alpha\beta}$. $\xi$ is the supersymmetric parameter, a $Spin(2)$ spinor, $\lambda$ contains the fermionic variables of the multiplet in a $Spin(2)$ spinor, and $x^\mu$ are four real bosonic variables. $q$ labels the multiplet. Dimensional reduction of this transformation recovers the transformations of the $\mathcal N = (2,2)$ vector multiplet, presented in \cite{Sonner:2008fi}. Writing the spinor indices explicitly in equation \eqref{eqeqeq},
\begin{equation}
\delta ( \lambda^q )_\alpha = i\dot x^\mu_q \sigma^4_{\alpha\rho}\left(\bar \sigma^\mu\right)^{\rho\beta} \xi_\beta \;\;,
\end{equation}
where $\sigma^4$, related to the identity, was taken into account to give the right spinoral index structure to the equation. From now on, $\sigma^4$ matrices will not be left in the equations.

The most general connection in the many multiplet case is
\begin{equation}\label{outroLag}
{\cal A} = A_\mu^q (x)\dot x_\mu^q + (B_{pq}(x)\lambda^p \lambda^q + \text{h.c.}) + K_\mu^{qr} (x)\bar\lambda^q \bar \sigma^\mu \lambda^r \;\; .
\end{equation}

Comparing this transformations with the vector multiplet transformations under $\mathcal N = (2,2)$ supersymmetry \cite{Sonner:2008fi}, we see that they are the same, with the identification $\dot x^4_q = D^q$. At the level of the connection ${\cal A}$, the identifications to be made are $A_4^p=-H_p$, $i K_4^{pq} = C_{pq}$ and $\vec K^{pq}= \vec C_{pq} $.

Doing variations to the connection \eqref{outroLag}, we obtain
\begin{equation}
\begin{split}
\delta {\cal A} =& \frac{d}{dt}\left( A_\mu^q \delta x_\mu^q \right) + \left( \partial_\nu^s A_\mu^q - \partial_\mu^q A_\nu^s \right)\left(i \bar\lambda^r \bar \sigma^\nu \xi - i \bar \xi \sigma^\nu \lambda^r   \right)\dot x_\mu^q \\
&+\partial_\nu^s K_\mu^{qr} \left( i \bar \lambda^s \bar \sigma^\nu \xi - i \bar \xi \sigma^\nu \lambda^s \right) \bar \lambda ^q \bar \sigma^\mu \lambda^r + K_\nu^{qr} \bar \xi \sigma^\mu \bar \sigma^\nu \lambda^r \dot x_\mu^q + K_\nu^{rq} \bar \lambda^r \bar \sigma^\nu \bar \sigma^\mu \xi \dot x^q_\mu \\
&+\left[ \partial_\mu^r B^{pq} \left( i \bar \lambda^r \bar \sigma^\mu \xi - i \bar \xi \sigma^\mu \lambda^r \right) \lambda^p \lambda^q  - B^{pq} \dot x_\mu^p \xi \sigma^\mu \lambda^q + B^{pq} \lambda^p \dot x_\mu^q \bar \sigma^\mu \xi + h.c. \right] \;\;.
\end{split}
\end{equation}
In order for the transformation \eqref{susytrans} to be a symmetry, equation \eqref{sym} must hold:
\begin{equation}
\begin{split}
\delta {\cal A} =& \frac{d}{dt}\left( A_\mu^q \delta x_\mu^q \right) + i\left[A_\mu^q , A_\nu^r \right] \dot x_\mu^q \left( i \bar \lambda^r \bar \sigma^\nu \xi - i \bar \xi \sigma^\nu \lambda^r\right) \\
&+i\left[B^{pq} , A_\nu^r \right] \lambda^p \lambda^q \left( i \bar \lambda^r \bar \sigma^\nu \xi - i \bar \xi \sigma^\nu \lambda^r\right)\\
&+i\left[(B^{pq})^\dagger , A_\nu^r \right] \bar\lambda^q \bar\lambda^p \left( i \bar \lambda^r \bar \sigma^\nu \xi - i \bar \xi \sigma^\nu \lambda^r\right)\\
&+i\left[K_\mu^{qs}, A_\nu^r \right] \bar\lambda^q \bar\sigma^\mu\lambda^s \left( i \bar \lambda^r \bar \sigma^\nu \xi - i \bar \xi \sigma^\nu \lambda^r\right) \;\;.
\end{split}
\end{equation}
These two equations have to be equal. From the coefficient of $\xi \lambda$, we conclude $B^{pq}=0$. From the coefficient of $\dot x \bar \lambda \xi$, we get
\begin{equation}\label{eq40}
F_{\nu\mu}^{rq} \dot x^\mu_q i \bar \lambda^r \bar \sigma^\nu \xi + K_\nu^{rq} \bar \lambda^r \bar \sigma^\nu \bar \sigma^\mu \xi \dot x^\mu_q = 0 \;\; .
\end{equation}
Finally, we also get the condition
\begin{equation}\label{eq41}
\mathcal D_\nu^s K_\mu^{qr} i \bar \lambda^q \bar \sigma^\mu \lambda^r \bar \lambda^s \bar \sigma^\nu \xi = 0 \;\;.
\end{equation}
Since $K$ is related with $F$ through equation \eqref{eq40}, it inherits some symmetry properties from $F$, including conditions coming from the Bianchi identity. After long and tedious calculations, it is possible to show that \eqref{eq41} is trivially satisfied. Hence, this equation does not impose any further condition on $K$.

Now, we are going to extract more information from \eqref{eq40}. We know that
\begin{equation}
\bar\sigma^\nu \bar\sigma^\mu= \left\{
\begin{array}{ll}
\delta^{ij} + i \epsilon_{ijk}\sigma^k & \text{  if   } \nu=i,\,\mu=j \;,\\
i\sigma^i & \text{  if   } \nu=i,\,\mu=4\;,\\
i\sigma^j & \text{  if   } \nu=4,\,\mu=j\;,\\
-1 & \text{  if   } \nu=4,\,\mu=4\;,
\end{array} \right.
\end{equation}
so, equation \eqref{eq40} is telling us
\begin{subequations}
\begin{align}
-F_{4j}^{rq}+K_j^{rq} =& 0 \;\; ,\\
-F_{44}^{rq}-K_4^{rq} =& 0 \;\; ,\\
F_{kj}^{rq} + \epsilon_{ijk} K_i^{rq} + \delta_{kj} K_4^{rq} =& 0 \;\; ,\\
F_{k4}^{rq} + K_k^{rq} = 0 \;\; .
\end{align}
\end{subequations}
These equations contain
\begin{equation}
F_{\mu\nu}^{rq} = -\frac{1}{2} \epsilon_{\mu\nu\rho\sigma} F_{\rho\sigma}^{rq} - \delta_{\mu\nu} K_4^{rq} \;\; .
\end{equation}
This is the same generalisation of the self-dual equation as the one obtained before \eqref{ginst}.

\bibliographystyle{is-unsrt}
\bibliography{PaperSubmission}
\end{document}